\documentclass[sn-nature]{sn-jnl}

\newcommand{\aj}{Astron. J.}

\newcommand{\araa}{Annual Review of Astronomy and Astrophysics}
\newcommand{\apj}{Astrophys. J.}
\newcommand{\apjl}{Astrophys. J., Lett.}

\newcommand{\apss}{Astrophysics and Space Science}
\newcommand{\aap}{Astron. Astrophys.}

\newcommand{\icarus}{Icarus}

\newcommand{\mnras}{Mon. Not. R. Astron. Soc.}

\newcommand{\pasp}{Publ. Astron. Soc. Pacific}
\newcommand{\pasj}{Publications of the Astronomical Society of Japan}

\newcommand{\nat}{Nature}

\usepackage{graphicx}%
\usepackage{multirow}%
\usepackage{amsmath,amssymb,amsfonts}%
\usepackage{amsthm}%
\usepackage{mathrsfs}%
\usepackage[title]{appendix}%
\usepackage{xcolor}%
\usepackage{textcomp}%
\usepackage{manyfoot}%
\usepackage{booktabs}%
\usepackage{algorithm}%
\usepackage{algorithmicx}%
\usepackage{algpseudocode}%
\usepackage{listings}%
\usepackage{caption}
\usepackage{setspace}

\theoremstyle{thmstyleone}%
\theoremstyle{thmstyletwo}%

\theoremstyle{thmstylethree}%

\begin{document}

\title[Observationally derived magnetic field strength and 3D components in the HD 142527 disk]{\bf Observationally derived magnetic field strength and 3D components in the HD 142527 disk}


\author*[1,2]{\fnm{Satoshi} \sur{Ohashi}}\email{satoshi.ohashi@nao.ac.jp}

\author[3]{\fnm{Takayuki} \sur{Muto}}

\author[4]{\fnm{Yusuke} \sur{Tsukamoto}}
\author[1,5]{\fnm{Akimasa} \sur{Kataoka}}
\author[6]{\fnm{Takashi} \sur{Tsukagoshi}}
\author[7]{\fnm{Munetake} \sur{Momose}}
\author[1]{\fnm{Misato} \sur{Fukagawa}}
\author[2]{\fnm{Nami} \sur{Sakai}}

\affil*[1]{\orgdiv{ALMA Project}, \orgname{National Astronomical Observatory of Japan}, \orgaddress{\street{2-21-1 Osawa}, \city{Mitaka}, \postcode{181-8588}, \state{Tokyo}, \country{Japan}}}

\affil[2]{\orgdiv{Cluster for Pioneering Research}, \orgname{RIKEN}, \orgaddress{\street{2-1 Hirosawa}, \city{Wako-shi}, \postcode{351-0198}, \state{Saitama}, \country{Japan}}}

\affil[3]{\orgdiv{Division of Liberal Arts}, \orgname{Kogakuin University}, \orgaddress{\street{1-24-2 Nishi-Shinjyuku}, \city{Shinjyuku-ku}, \postcode{163-8677}, \state{Tokyo}, \country{Japan}}}
\affil[4]{\orgdiv{Graduate Schools of Science and Engineering}, \orgname{Kagoshima University}, \orgaddress{\street{1-21-35 Korimoto}, \city{Kagoshima}, \postcode{890-0065}, \state{Kagoshima}, \country{Japan}}}

\affil[5]{\orgdiv{Department of Astronomical Science}, \orgname{University for Advanced Studies (SOKENDAI)}, \orgaddress{\street{2-21-1 Osawa}, \city{Mitaka}, \postcode{181-8588}, \state{Tokyo}, \country{Japan}}}

\affil[6]{\orgdiv{Faculty of Engineering}, \orgname{Ashikaga University}, \orgaddress{\street{268-1 Ohmae-cho}, \city{Ashikaga}, \postcode{326-8558}, \state{Tochigi}, \country{Japan}}}

\affil[7]{\orgdiv{College of Science}, \orgname{Ibaraki University}, \orgaddress{\street{2-1-1 Bunkyo}, \city{Mito}, \postcode{310-8512}, \state{Ibaraki}, \country{Japan}}}


\abstract{\bf In protoplanetary disks around young stars, magnetic fields play an important role for disk evolution and planet formation. Polarized thermal emission from magnetically aligned grains is one of the reliable methods to trace magnetic fields. However, it has been difficult to observe magnetic fields from dust polarization in protoplanetary disks because other polarization mechanisms involving grown dust grains become efficient. Here, we report multi-wavelength (0.87 mm, 1.3 mm, 2.1 mm, and 2.7 mm) observations of polarized thermal emission in the protoplanetary disk around HD 142527, showing the lopsided dust distribution. We revealed that the smaller dust still exhibits magnetic alignment in the southern part of the disk. Furthermore, angular offsets between the observed magnetic field and the disk azimuthal direction were discovered, which can be used as a method to measure the relative strengths of each component (radial ($B_r$), azimuthal ($B_\phi$), and vertical ($B_z$)) of 3D magnetic field. Applying this method, we derived the magnetic field around a 200-au radius from the protostar as $|B_r |:|B_\phi |:|B_z | \sim 0.26:1:0.23$ and a strength of $\sim 0.3$ milli-Gauss. Our observations provide some key parameters of magnetic activities including the plasma beta, which have only been assumed in theoretical studies. In addition, the radial and vertical angular momentum transfer are found to be comparable, which poses a challenge to theoretical studies of protoplanetary disks.}

\maketitle


HD 142527 is a Herbig Fe star \cite{1996A&A...315L.245W} with a mass and age of $\sim 2.4$ $M_\odot$ and $\sim3$ Myr, respectively \cite{2019AJ....157..159A,2006ApJ...636L.153F,2011A&A...528A..91V}, and is located 157 pc away from our solar system \cite{2016A&A...595A...2G,2016A&A...595A...1G}. The accretion rate to the star is estimated to be $10^{-7}$  $M_\odot$  yr$^{-1}$ \cite{2014ApJ...790...21M}. The disk surrounding the young star exhibits a wide gap with a radial width of about 100 au and a horseshoe-shaped ring structure \cite{2013Natur.493..191C,2013PASJ...65L..14F}. This high-contrast emission on north side brighter than south suggests that larger dust grains exist in the northern region, while smaller dust grains are located in the southern region \cite{2018ApJ...864...81O,2019PASJ...71..124S}. Therefore, the protoplanetary disk around HD 142527 is an ideal laboratory to investigate the grain growth, which is considered to be the first step in planet formation.
We analyzed the polarization data of the dust continuum emission at wavelengths of 0.87 mm (Band 7), 1.3 mm (Band 6), 2.1 mm (Band 4), and 2.7 mm (Band 3) obtained by the Atacama Large Millimeter/submillimeter Array (ALMA) (see Methods).

\section*{Results}

Figure 1 shows images of Stokes {\it I}, polarized intensity (PI), and polarization fraction (PF) at these wavelengths. Polarization vectors (PA) are also plotted on the PI and PF images. The horseshoe structure can be recognized in the Stokes {\it I} images at the all wavelengths as the previous observations revealed \cite{2013Natur.493..191C,2013PASJ...65L..14F,2018ApJ...864...81O,2019PASJ...71..124S}. Although the Stokes {\it I} images are similar in these wavelengths, the polarization images show different morphologies. In particular, the polarization vectors and fraction in the northern region significantly change with the wavelength. The 0.87 mm polarization vectors show a flip around the peak of the Stokes {\it I} emission as shown in the previous studies \cite{2018ApJ...864...81O,2016ApJ...831L..12K}, while the 2.1 mm and 2.7 mm polarization vectors appear to be almost azimuthally aligned. The 1.3 mm polarization vectors also show the flip, although less prominent \cite{2020ApJ...901...71S}. These different polarization patterns can be explained by the contribution of the self-scattering, which is efficient only when the maximum grain size is similar to the observing wavelength. Therefore, the dust grains of $\approx 100$ $\mu$m are abundant in the northern region \cite{2015ApJ...809...78K,2016ApJ...831L..12K}. In contrast to the northern region, the southern region shows almost the same polarization pattern regardless of the observed wavelength. All the polarization vectors are in the radial direction, and the polarization fraction reaches as high as 15 \%.

\begin{figure}[h]
\centering
\includegraphics[width=12.cm,bb=0 0 511 372]{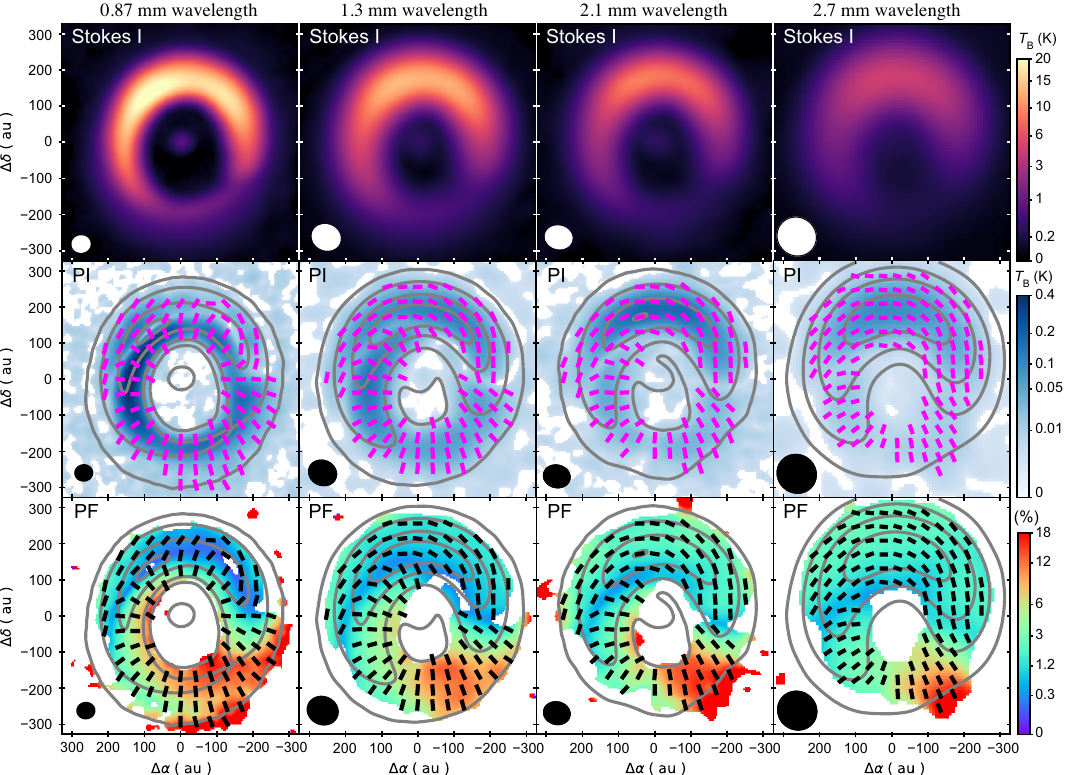}
\caption{{\bf Polarized thermal emission from dust grains of the protoplanetary disk around HD 142527. $\mid$} Stokes {\it I} (top), polarized intensity (PI) (middle), and polarization fraction (PF) (bottom) at 0.87 mm, 1.3 mm, 2.1 mm, and 2.7 mm wavelengths are shown, respectively. Polarization vectors are also shown on the PI and PF images. The PF and polarization vectors are plotted where the polarized intensity is detected above the 3$\sigma$ noise level. Contours on the PI and PF images indicate the brightness temperatures of the Stokes {\it I} emission of [0.1, 1, 10, 20] K for 0.87 mm, [0.1, 1, 5, 10, 20] K for 1.3 mm, [0.1, 1, 5, 9] K for 2.1 mm, and [0.06, 0.3, 1, 3] K for 2.7 mm. The resolution (beam size) is shown as a small ellipse in the lower left corner of each panel.}\label{fig1}
\end{figure}

\begin{figure}[h]
\centering
\includegraphics[width=12.cm,bb=0 0 250 207]{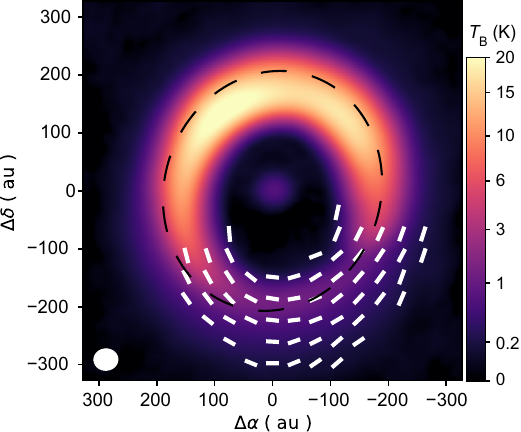}
\caption{{\bf The magnetic field morphology of the HD 142527 disk $\mid$} The magnetic field is shown by the white segments superimposed on the 0.87 mm Stokes {\it I} continuum image, which is derived by rotating the 0.87 mm polarization vectors by 90 degrees because the dust polarization is caused by the thermal emission of magnetically aligned dust grains in the southern region. The black dashed line indicates the azimuthal direction of the disk with a radius of 210 au.
}\label{fig2}
\end{figure}

The consistent polarization vectors and polarization fractions in the southern region (position angle of $100-250$ degrees, see Methods and Supplementary Figures 1, 2, and 3) suggest that the polarization is produced by grain alignment rather than self-scattering. High ($\sim15$\%) polarization in radial direction is expected to originate from the magnetically aligned grains \cite{2007MNRAS.378..910L,2015ARA&A..53..501A} under the toroidal magnetic field \cite{2022AJ....164..248H,2007ApJ...669.1085C,2017MNRAS.464L..61B}. In contrast, other alignment mechanisms, such as radiative alignment or mechanical alignment, predict azimuthally polarized emission \cite{2017ApJ...839...56T,2019ApJ...874L...6K,2019MNRAS.483.2371Y}, which may be the case for the 1.3 mm, 2.1 mm, and 2.7 mm polarizations in the northern region. It should be noted that the magnetic alignment also predicts azimuthally polarized emission at a millimeter wavelength when the dust size is larger than 1 mm due to a negative polarization fraction in the Mie regime \cite{2020A&A...634L..15G}. Therefore, we conclude that at least the smaller dust grains in the southern region are magnetically aligned, as in the similar case of the interstellar medium (ISMs) \cite{2007MNRAS.378..910L,2015ARA&A..53..501A}. This may imply that the dust grains in the southern region contain large iron inclusions to be aligned with magnetic field in protoplanetary disks \cite{2018ApJ...852..129H}. 

For the southern region, the polarization vector is expected to be perpendicular to the magnetic field direction projected onto the sky plane. Thus we can investigate the magnetic morphology by rotating the polarization vectors 90 degrees. Figure 2 indicates that the toroidal field is dominant, as expected from MHD simulations \cite{2007Ap&SS.311...35W,2013ApJ...769...76B,2014ApJ...784..121S}. To further constrain the magnetic field structure, we measured the angular offset of the observed magnetic field from the purely toroidal case, assuming the disk inclination and position angle of 27$^\circ$ and 161$^\circ$, respectively \cite{2016A&A...595A...1G,2021A&A...650A..59B}.

\section*{Three-Dimensional Magnetic Field}

Figure 3 shows the angular difference ($\Delta\phi=B_\phi-B_{\rm obs}$) between the observed magnetic field ($B_{\rm obs}$) and the pure toroidal magnetic field ($B_\phi$) as a function of a position angle at $200\pm50$ au from the central star in the southern region. From this figure, we found that the observed magnetic field has systematic offsets of about 10 degrees from the toroidal case. We interpret these offsets as being caused by the radial and vertical components of the magnetic fields in the disk.

\begin{figure}[h!]
\centering
\includegraphics[width=12.cm,bb=0 0 250 188]{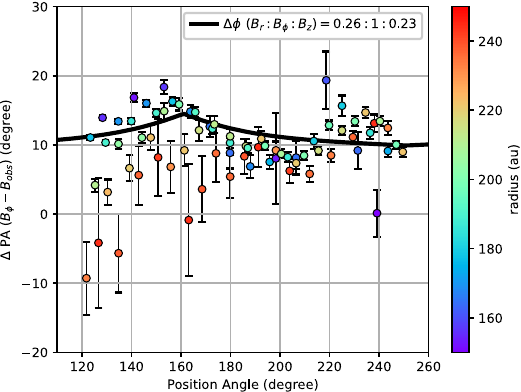}
\caption{{\bf The angular difference between the observed magnetic field and the toroidal magnetic field as a function of a position angle. $\mid$} We plot the angular difference ($\Delta\phi=B_\phi-B_{\rm obs}$) between the observed magnetic field ($B_{\rm obs}$) and the pure toroidal magnetic field ($B_\phi$) as a function of a position angle at $200\pm50$ au from the central star in the southern region. The color indicates the radius from the central star. The data were obtained using Nyquist sampling (a half of beam size) with a disk radius of $200\pm50$ au. The data is presented as the median of the distribution with the errors corresponding to the 68th percentiles. The systemic offset of abound 10 degrees can be found. This offset is caused by the contributions of the radial ($B_r$) and vertical ($B_z$) magnetic fields. The black line shows the best fit parameter of $|B_r |:|B_\phi |:|B_z |=0.26:1: 0.23$ to explain this offset. Note that the fitting errors are not shown because they are not meaningful due to the large reduced chi-square value of approximately 9 for the best-fit model.}\label{fig3}
\end{figure}

\subsection*{Single Layer Model}

Here, we propose a method to estimate the relative strengths of the three-dimensional components of the magnetic fields, radial ($B_r$), azimuthal ($B_\phi$), and vertical ($B_z$). The $B_r$ and $B_z$ components act to deviate the sky-projected magnetic field vector from the purely toroidal direction.  Their effects vary at different locations on the disk (see Figure 4), causing azimuthal variations in the angular offsets even if the intrinsic magnetic field exerted on the disk is homogeneous. The angular offset is estimated as
\begin{equation}
\Delta\phi=\arctan\Big\{\frac{  (B_r)_{\phi\perp} + (B_{z,proj})_{\phi\perp}  }{  B_\phi  + (B_r)_{\phi} + (B_{z,proj})_{\phi}   } \Big\}\label{eq1}
\end{equation}
where $(B_r)_{\phi\perp}$ is the component perpendicular to the azimuthal direction in sky-projected $B_r$, $(B_r)_{\phi}$ is the component parallel to the azimuthal direction in sky-projected $B_r$, $B_{z,proj}=B_z \sin{\rm (inclination)}$ is the sky-projected strength, $(B_{z,proj})_{\phi\perp}$ is the component perpendicular to the azimuthal direction in $B_{z,proj}$, $(B_{z,proj})_{\phi}$ is the component parallel to the azimuthal direction in $B_{z,proj}$.  These parameters are functions of the position angle by considering the projected direction of the grain alignment in the inclined disk geometry. Thus, the variation of the offset value, $\Delta\phi$, along the position angle allows us to estimate the relative strengths of the three-dimensional magnetic field. We find that $|B_r |:|B_\phi |:|B_z |  = 0.26:1:0.23$ best fits the observations from the chi-squared fit as shown in Figure 3. The reduced chi-square value for the best-fit model is approximately 9, suggesting that more detailed observations and refined models would be necessary to understand fine magnetic structure such as the radial distribution. Nevertheless, as a whole view, the absolute values and variations of $\Delta\phi$ suggest that the azimuthal (toroidal) magnetic field is dominant, while the radial and vertical magnetic field strengths are almost comparable. Note that there is uncertainty in the magnetic field direction because the polarization vectors only indicate the alignment direction. Therefore, we used the absolute values for the relative strength.

We discuss the dependence of the relative strengths of $|B_r |:|B_\phi |:|B_z |$ on the angular offsets ($\Delta\phi$) by changing these parameters and confirm the robustness of the best-fit results.
Figure 5 shows four different cases of the relative strengths of the magnetic field. The black line indicates the best-fit parameter, and the dashed color lines indicate the different parameters of the relative strengths as $|B_r |:|B_\phi |:|B_z |  =0.3:1:0.1$ (red), $0.25:1:0.5$ (blue), and $0.2:1:0.3$ (green), respectively. As shown in this figure, the systematic offset of $\Delta\phi$ is sensitive to $B_r$, while the variation of $\Delta\phi$ along the position angle is sensitive to $B_z$. Therefore, the absolute values of $\Delta\phi$ and its dependence on the position angle allow one to derive the relative strengths of  $B_r$, $B_\phi$, and $B_z$. By comparing these different parameters, we suggest that the best fit parameter of  $|B_r |:|B_\phi |:|B_z |  = 0.26:1:0.23$ would be a reasonable estimate.

\begin{figure}[h!]
\centering
\includegraphics[width=12.cm,bb=0 0 251 262]{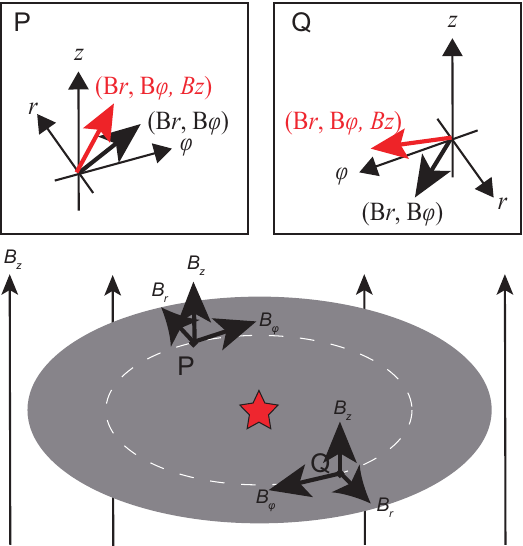}
\caption{{\bf A schematic view of the three-dimensional magnetic field structure. $\mid$} We propose that the dependence of the angular offsets ($\Delta\phi$) on the disk position angle (shown as Figure 3) is caused by the three-dimensional magnetic field structure of the disk as $B_r$, $B_\phi$, and $B_z$. In the position P, the radial and vertical magnetic fields point in the same direction, strengthening the offset from the azimuthal direction, while in position Q, the radial and vertical magnetic fields point in the opposite direction, weakening the offset from the azimuthal direction. Therefore, $\Delta\phi$ will vary with the disk position angle.}\label{fig4}
\end{figure}

\begin{figure}[h!]
\centering
\includegraphics[width=12.cm,bb=0 0 250 188]{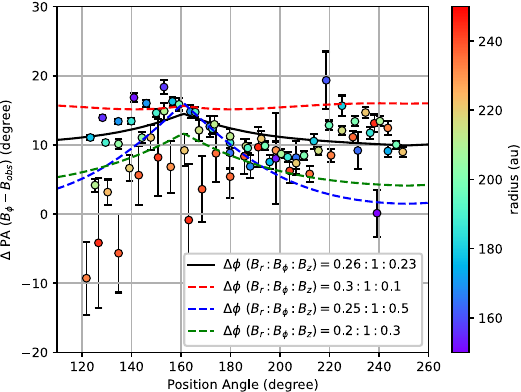}
\caption{{\bf The angle offset with various relative magnetic field strength. $\mid$} We plot the angular difference ($\Delta\phi=B_\phi-B_{\rm obs}$) between the observed magnetic field ($B_{\rm obs}$) and the pure toroidal magnetic field ($B_\phi$) as a function of a position angle at $200\pm50$ au from the central star in the southern region. The color indicates the radius from the central star. The data were obtained using Nyquist sampling (a half of beam size) with a disk radius of $200\pm50$ au.  The data is presented as the median of the distribution with the errors corresponding to the 68th percentiles. The systemic offset of abound 10 degrees can be found. This offset is caused by the contributions of the radial ($B_r$) and vertical ($B_z$) magnetic fields. The black line shows the best fit parameter of $|B_r |:|B_\phi |:|B_z |=0.26:1: 0.23$ to explain this offset. The dashed color lines indicate the different parameters of the relative magnetic field strengths as $|B_r |:|B_\phi |:|B_z |=0.3:1:0.1$ (red), $0.25:1:0.5$ (blue), and $0.2:1:0.3$ (green), respectively to show the dependence of the $\Delta\phi$ on the relative magnetic field strengths.}\label{fig5}
\end{figure}

\subsection*{Dual Layer Model}

It can be expected that the $B_\phi$ and $B_r$ are likely to be in opposite directions with respect to the disk midplane due to the entrainment of the initial $B_z$ field by the disk rotation (Extended Data Figure 1). 
Thus, the observed offset value may be the combination of the upper disk layer ($\Delta\phi_+$) and the lower disk layer ($\Delta\phi_-$) (dual layer model). By considering this effect, we fitted the offset values as a function of the position angle (see Methods). 
The best fit resulted in $|B_r |:|B_\phi |:|B_z |  =0.21:1:<0.1$ as shown by the black line in Figure 6. The reduced chi-square value for the best-fit model is approximately 16, which is larger than that of the single layer model. This is because the variation of $\overline{\Delta\phi}$ (a combination of $\Delta\phi_{+}$ and $\Delta\phi_{-}$, see Methods Equation \ref{eq2}) along the position angle cannot be reproduced by the dual layer model. To confirm this, the different parameters of the magnetic field strengths are also plotted in the colored dashed lines. We found that larger $B_z$ causes smaller angular offset of $\overline{\Delta\phi}$ around the disk major axis (position angle of 150 degrees) and larger angular offset around the disk minor axis (position angle of 220 degrees). This is because $B_z$ has the same direction with $B_\phi$ in the disk major axis (suppress $\overline{\Delta\phi}$) and with $B_r$ in the disk minor axis (enhance $\overline{\Delta\phi}$). We also found that $B_z$ becomes more efficient when it has the same direction with $B_\phi$ than the opposite direction to $B_\phi$ in the dual layer model because the projected length of the elongated dust grains becomes longer. In contrast, the previous single layer model can enhance $\Delta\phi$ in the disk major axis and suppress  $\Delta\phi$ in the disk minor axis by changing the direction of $B_z$ with respect to $B_r$. Therefore, the best fit parameter of the single layer model suggests that the $B_r$ and $B_z$ directions are opposite to each other, i.e., if $B_r$ is outward, then $B_z$ is from south to north direction in the sky-projected direction.

\begin{figure}[h!]
\centering
\includegraphics[width=12.cm,bb=0 0 250 141]{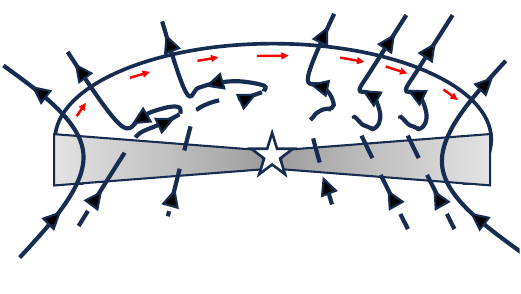}
 \captionsetup{labelformat=empty,labelsep=none}
\caption{{\bf Extended Data Figure 1 $\mid$ A schematic view of the 3D configuration of the magnetic field in a protoplanetary disk.} The black lines show the magnetic field, and the red arrows indicate the rotation direction. $B_\phi$ and $B_r$ are generated by the rotation of the initial state, $B_z$. Then, $B_\phi$ and $B_r$ are likely to be in opposite directions with respect to the disk midplane due to the entrainment of the initial $B_z$ field by the disk rotation.}\label{fig9}
\end{figure}

These models imply that the single layer model can reproduce the observed features of the systematic offset and variation along the position angle better than the dual layer model. This suggests that the magnetic field across the disk midplane is not purely symmetric due to turbulence. Some MHD simulations have shown such asymmetric magnetic field structures \cite{2017ApJ...845...75B}, although few simulations have been performed for disks with substructures such as rings and crescents \cite{2018MNRAS.477.1239S}. It is possible that the magnetic field in a crescent disk like this one is not a simple symmetric structure across the midplane. It will be necessary to investigate further simulations for the magnetic field in ring or crescent structures. Note that the emission is estimated to be optically thin ($\tau\sim 0.1$) in the southern region at a wavelength of 0.87 mm, suggesting that the polarized emission would be the averaged value along the vertical direction of the scale height and would mainly trace the midplane region. This also supports the preference of the single layer model over the dual layer model, because $B_\phi$ is supposed to be dropped at the disk midplane for the dual layer model to flip the direction, while the observations indicate that $B_\phi$ is the dominant component of the disk magnetic field. Although different values of $B_z$ were estimated by the two models, $B_r$ shows almost the same strength of $B_r\approx (0.2-0.3)B_\phi$. This is because $B_r$ is mainly determined by the systematic offset of $\Delta\phi$, while $B_z$ is determined by the variations of $\Delta\phi$ with the position angle. Therefore, $B_r$ can be estimated better than $B_z$, and $B_r\approx(0.2-0.3)B_{\phi}$ would be robust. Since the reduced chi-square values for both the single and dual layer models are much larger than 1 (9 for the single layer model and 16 for the dual layer model), more detailed observations and refined models would be necessary to understand fine magnetic structure such as the radial distribution of $B_r$ and $B_z$.

\begin{figure}[h!]
\centering
\includegraphics[width=12.cm,bb=0 0 250 188]{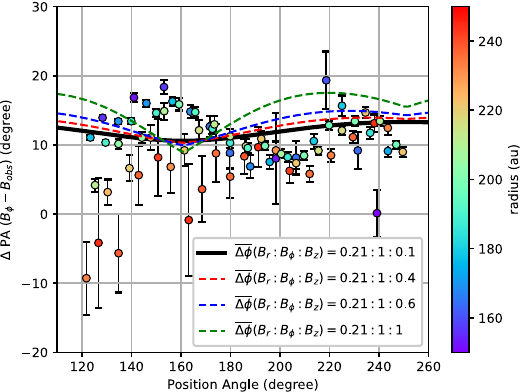}
\caption{
{\bf The angle differences with various relative magnetic field strengths of the dual layer model. $\mid$} We plot the angular difference ($\Delta\phi=B_\phi-B_{\rm obs}$) between the observed magnetic field ($B_{\rm obs}$) and the pure toroidal magnetic field ($B_\phi$) as a function of a position angle at $200\pm50$ au from the central star in the southern region. The color indicates the radius from the central star. The data were obtained using Nyquist sampling (a half of beam size) with a disk radius of $200\pm50$ au. The data is presented as the median of the distribution with the errors corresponding to the 68th percentiles. The systemic offset of abound 10 degrees can be found. This offset is caused by the contributions of the radial ($B_r$) and vertical ($B_z$) magnetic fields. 
For the case of the dual layer model for the magnetic field, the best fit gives $|B_r |:|B_\phi |:|B_z |  \approx 0.21:1:<0.1$ as shown by the black line. The different parameters of the relative strengths are also plotted as the colored dashed lines as $|B_r |:|B_\phi |:|B_z |  =0.21:1:0.4$ (red), $0.21:1:0.6$ (blue), and $0.21:1:1$ (green), respectively.
}\label{fig6}
\end{figure}

\section*{Magnetic Field Strength and Activity}

Our results on the relative magnetic field strength allow us to estimate important physical parameters related to the disk magnetic field based on observations. To realize $B_z/B_\phi \sim 0.23$ in the disk around HD 142527, we estimated the ambipolar resistivity $\eta_{\rm A}\sim 3.0\times10^{17}$ cm$^2$  s$^{-1}$, a magnetic field strength of $B \sim 0.3$ milli Gauss (mG), the ambipolar Elsasser number Am of $\sim 0.4$, and the plasma beta of $\beta \sim 2.0\times10^2$. (see Methods). The Am parameter is the dimensionless number that characterizes the ambipolar diffusion effect. From these estimates, we suggest that the magnetorotational instability (MRI) can occur in the southern region of HD 142527, while Am and $\beta$ are close to the boundary between MRI-active and inactive parameters \cite{2011ApJ...736..144B}. This is consistent with near-infrared observations suggesting that the HD142527 disk is turbulent \cite{2021ApJ...921..173T}. Furthermore, the magnetic field strength of $0.3$ mG at $\sim 200$ au from the central star is consistent with a prediction from numerical simulations \cite{2015ApJ...798...84B} and within the upper limit constrained by the Zeeman observations \cite{2019A&A...624L...7V}. 
By assuming that $B_z$ has a power-law profile of $B_z\propto r^p$, the power-law index, $p$, can be estimated to be $p\sim -1$ from the relative strengths of $B_r/B_z$  (see Methods). Such a power-law profile of the vertical magnetic field suggests that the magnetic field increases moderately toward the center. The fact that the power exponent of the magnetic field is flatter than 2 suggests that the magnetic flux freezing has thawed and that both magnetic field advection and outward diffusion play a role \cite{2014MNRAS.441..852G,2024PASJ...76..674T}.

The relative strengths of the three components of the magnetic field also have important implications for disk evolution processes \cite{2016ARA&A..54..135H,2007Ap&SS.311...35W,2013ApJ...769...76B}. For example, the angular momentum transport due to the magnetic field can be inferred from the Maxwell's stress, $M$, as $M_{r\phi}=\frac{B_r B_\phi}{4\pi}$ and $M_{z\phi}=\frac{B_z B_\phi}{4\pi}$ along the radial (mass accretion) and vertical (disk wind) directions, respectively. The alpha value is commonly used to evaluate the angular momentum transport as $\alpha_{r\phi}=\frac{M_{r\phi}}{\rho c_{\rm s}^2}=\frac{2B_r}{\beta_\phi B_\phi}$ and $\alpha_{z\phi}=\frac{M_{z\phi}}{(\rho c_{\rm s}^2}=\frac{2B_z}{\beta_\phi B_\phi}$. From our analysis, $\beta$ was found to be $\sim~ 2.0\times10^2$. Therefore, the alpha values are $\alpha_{r\phi} \sim \frac{0.53}{\beta}=2.5\times10^{-3}$ and $\alpha_{z\phi}  \sim  \frac{0.6}{\beta}  = 2.2\times10^{-3}$, respectively. We therefore conclude that the angular momentum transports in the radial and vertical directions are comparable in the $\sim 200$ au region of the protoplanetary disk around HD 142527.

In this paper, we demonstrated that the dust polarization can still be used as a magnetic tracer for protoplanetary disks when the dust grains are as small as the ISMs. The three-dimensional magnetic field can be probed for a disk with moderate inclination. The analyses allow us to estimate several physical parameters related to the magnetic field in the disk and provide insight into the angular momentum transfer, which is an important process determining the evolution of the disk. This method will also be applicable to other disks. Thus, our observations shed light on the study of the magnetic field in protoplanetary disks.

\newpage

\section*{Methods}
\setcounter{figure}{0}
\subsection*{Observations and data reduction}

We used the archival data of the ALMA projects 2015.1.00425.S, 2017.1.00987.S, 2018.1.01172.S, and 2022.1.00406.S for this study. The results of the projects 2015.1.00425.S and 2018.1.01172.S have been presented previously \cite{2018ApJ...864...81O,2016ApJ...831L..12K,2020ApJ...901...71S}. These projects observed HD 142527 in Full Stokes polarization. The basic parameters of these observations are summarized in Supplementary Table 1. The polarization calibrator was observed $3-4$ times with $\sim6$ minutes of integration time during each observing schedule to calibrate the instrumental polarization (D-terms), cross-hand delay, and cross-hand phase. The reduction and calibration of the data were performed in a standard manner using CASA \cite{2007ASPC..376..127M}.

Project 2015.1.00425.S were carried out on 11 March 2016 during its Cycle 3 operation and on 21 May 2017 during its Cycle 4 operation. The correlator setup consists of four spectral windows with a bandwidth of 1.75 GHz centered at sky frequencies of 336.5, 338.5, 348.5, and 350.5 GHz, providing a total bandwidth of $\sim 7.5$ GHz. The total integration times for the target were 73 minutes in Cycle 3 operation and 80 minutes in Cycle 4 operation.

Project 2017.1.00987.S were carried out on 7 January 2018 during its Cycle 5 operation. The correlator setup consists of four 2 GHz spectral windows centered at sky frequencies of 137.995, 139.932, 149.995, and 151.995 GHz, providing a total bandwidth of $\sim 8$ GHz.

Project 2018.1.01172.S were carried out on 29 April 2019 during its Cycle 6 operation. The correlator setup consists of four spectral windows, and one band was set for 1.3 mm dust continuum emission centered at sky frequency of 234.5 GHz with a bandwidth of 2 GHz. The other three windows had a bandwidth of 59 MHz and were centered on CO ($J=2-1$), $^{13}$CO ($J=2-1$), and C$^{18}$O ($J=2-1$) emission lines. In this paper, we have used the spectral window of the 1.3 mm dust continuum emission.

2022.1.00406.S were carried out on 26, 27, 30, and 31 March 2023 during its Cycle 9 operation. The correlator setup consists of four spectral windows, and one band was set for 2.7 mm dust continuum emission centered at sky frequency of 112 GHz with a bandwidth of 2 GHz. The other three windows had a bandwidth of 59 MHz and were centered on $^{13}$CO ($J=1-0$), C$^{18}$O ($J=1-0$), and CN ($N=1-0, J=1/2 - 1/2, F=1/2 - 1/2$) emission lines. In this paper, we have used the spectral window of the 2.7 mm dust continuum emission.

\subsection*{Imaging}

Stokes {\it I}, {\it Q}, and {\it U} images of the calibrated visibility data were generated by the CASA task tclean. To improve image fidelity, we performed an iterative phase-only self-calibration using the initial CLEAN image as the first model image in CASA 6.5.1 for all projects expect 2022.1.00406.S, since 2022.1.00406.S was not improved by self-calibration. The interval time to solve the complex gain was reduced from inf, 300 s, and finally to 60 s. All images were generated with Briggs weighting. The robust parameters and the resulting parameters of the images are summarized in Supplementary Table 2. 
The PI was calculated from Stokes {\it Q} and {\it U} and has a positive bias because the Stokes {\it Q} and {\it U} components give the polarized intensity, $\sqrt{Q^2+U^2}$. This bias is a particularly noticeable in low signal-to-noise measurements. We therefore debiased the PI image as ${\rm PI}=\sqrt{Q^2+U^2-\sigma_{\rm PI}^2}$, where $\sigma_{\rm PI}$ is the rms noise level \cite{2006PASP..118.1340V,2015JAI.....450005H}.  The PF image was generated by $\rm PF (\%)=PI/(Stokes\  {\it I})\times100$, where the PI emission is above the noise level of 3$\sigma_{\rm PI}$. The polarization vectors were calculated from Stokes {\it Q} and {\it U} as ${\rm PA}=\frac{1}{2} \arctan(\frac{\rm Stokes\ {\it U}}{\rm Stokes\  {\it Q}})$ where the PI emission is above the 3$\sigma_{\rm PI}$ noise level.

\subsection*{Imaging of the differences of the polarization vectors and polarization fractions}

To avoid confounding by different spatial frequency components, the images of the polarization vector differences (Supplementary Figure 1) were generated using the same uv distance. For $\Delta {\rm PA}_{\rm 1.3mm-0.87mm}={\rm PA}_{\rm 1.3mm} - {\rm PA}_{\rm 0.87mm}$, the Stokes {\it Q} and {\it U} images of the 0.87 mm and 1.3 mm polarization data were generated by using a uv distance of $1.4\times10^4-5.5\times10^5$ $\lambda$. The Stokes {\it Q} and {\it U} images of the 0.87 mm data were smoothed to be a beam size of $0.517''\times0.452''$ with a position angle of $64.2^\circ$ to match the 1.3 mm data. The polarization vectors were then calculated from the Stokes {\it Q} and {\it U} images where the PI emission is above the 3sigma levels. The $\sigma_{\rm PI}$ is derived to be $35$ and $23$ $\mu$Jy beam$^{-1}$ for 0.87 mm and 1.3 mm, respectively.

For $\Delta {\rm PA_{2.1mm-0.87mm}=PA_{2.1mm} - PA_{0.87mm}}$, the Stokes {\it Q} and {\it U}  images of the 0.87 mm and 1.3 mm polarization data were generated by using a uv distance of $1.4\times10^4-1.2\times10^6$ $\lambda$. The Stokes {\it Q} and {\it U}  images of the 0.87 mm data were smoothed to be a beam size of $0.479''\times0.405''$ with a position angle of $76.1^\circ$ to match the 2.1 mm data.
Then, the polarization vectors were calculated from the Stokes {\it Q} and {\it U}  images. The polarization vectors were calculated where the PI emission is above the 3sigma levels, where $\sigma_{\rm PI}$ is derived to be $33$ and $7.6$ $\mu$Jy beam$^{-1}$ at 0.87 mm and 2.1 mm, respectively.
To make the images of the ratios of the polarization fractions (Supplementary Figure 3), the polarization fractions were also calculated by the above method. The polarization fractions were calculated where the PI emission is above the 3sigma levels.

For $\Delta {\rm PA_{2.7mm-0.87mm}=PA_{2.7mm} - PA_{0.87mm}}$, the Stokes {\it Q} and {\it U}  images of the 0.87 mm and 2.7 mm polarization data were generated by using a uv distance of $1.4\times10^4-4.7\times10^5$ $\lambda$. The Stokes {\it Q} and {\it U}  images of the 0.87 mm data were smoothed to be a beam size of $0.711''\times0.678''$ with a position angle of $50.6^\circ$ to match the 2.7 mm data.
Then, the polarization vectors were calculated from the Stokes {\it Q} and {\it U}  images. The polarization vectors were calculated where the PI emission is above the 3sigma levels, where $\sigma_{\rm PI}$ is derived to be $54$ and $7.4$ $\mu$Jy beam$^{-1}$ at 0.87 mm and 2.7 mm, respectively.
To make the images of the ratios of the polarization fractions (Supplementary Figure 3), the polarization fractions were also calculated by the above method. The polarization fractions were calculated where the PI emission is above the 3sigma levels.

To evaluate the wavelength dependence of the polarization patterns, we calculated the differences of the polarization vectors at 1.3 mm, 2.1 mm, and 2.7 mm with respect to the 0.87 mm polarization vectors by deriving $\Delta {\rm PA_{1.3mm-0.87mm}=PA_{1.3mm} - PA_{0.87mm}}$, $\Delta {\rm PA_{2.1mm-0.87mm}=PA_{2.1mm} - PA_{0.87mm}}$, and  $\Delta {\rm PA_{2.7mm-0.87mm}=PA_{2.7mm} - PA_{0.87mm}}$. Supplementary Figure 1 shows the images of the differences in polarization vectors. The northern region shows $\Delta {\rm PA_{1.3mm-0.87mm}}$, $\Delta {\rm PA_{2.1mm-0.87mm}}$, and $\Delta {\rm PA_{2.7mm-0.87mm}}$ of $\sim90$ degrees on the ridge of the horseshoe structure, indicating that the $\rm PA_{0.87mm}$ is perpendicular to $\rm PA_{1.3mm}$, $\rm PA_{2.1mm}$, and $\rm PA_{2.7mm}$. In contrast, the southern region shows that the polarization vectors are almost in the same direction ($\Delta {\rm PA_{1.3mm-0.87mm}}$, $\Delta {\rm PA_{2.1mm-0.87mm}}$, and $\Delta {\rm PA_{2.7mm-0.87mm}}\sim 0$ degree). Supplementary Figure 2 plots $\Delta {\rm PA_{1.3mm-0.87mm}}$, $\Delta {\rm PA_{2.1mm-0.87mm}}$, and $\Delta {\rm PA_{2.7mm-0.87mm}}$ as a function of a position angle of $110-250$ degrees on the ridge of the southern part of the horseshoe structure. The plotted data were taken from the pixels at the ridge position with a Nyquist sampling of 11 degrees for $\Delta {\rm PA_{1.3mm-0.87mm}}$, 10 degrees for $\Delta {\rm PA_{2.1mm-0.87mm}}$, and 15 degrees for $\Delta {\rm PA_{2.7mm-0.87mm}}$.  The differences of the polarization vectors are derived to be $\Delta {\rm PA_{1.3mm-0.87mm}}=3.9\pm0.6$ degrees, $\Delta {\rm PA_{2.1mm-0.87mm}}=1\pm1$ degrees, and $\Delta {\rm PA_{2.7mm-0.87mm}}=-5.2\pm3.4$ degrees. Supplementary Figures of 1 and 2 show that the polarization vectors have almost the same directions among the wavelengths. Although the larger angle offsets are seen at longer wavelength separations, this may be due to the weaker emission at longer wavelengths. The 2.7 mm polarization emission at Band 3 is the weakest emission with the largest beam size, leading to larger uncertainties in the polarization vectors due to the noise. In addition to the polarization vectors, Supplementary Figure 3 shows the ratios of the polarization fraction $\rm P_{frac,1.3 mm}/P_{frac,0.87 mm}$, $\rm P_{frac,2.1 mm}/P_{frac,0.87 mm}$, and $\rm P_{frac,2.7 mm}/P_{frac,0.87 mm}$ (method). In the northern region, the polarization fraction is significantly different, with $\rm P_{frac,1.3 mm}/P_{frac,0.87 mm}$ reaching values as high as $\sim5$ and $\rm P_{frac,2.1 mm}/P_{frac,0.87 mm}$ and $\rm P_{frac,2.7 mm}/P_{frac,0.87 mm}$ reaching values as high as $\sim10$. In contrast, in the southern region, the polarization fraction is almost the same value as all the polarization fraction ratios show $\sim1$.

\subsection*{The dual layers model for the angle offset}

In Figure 3, we have considered the single layer magnetic field to estimate the relative strengths. However, $B_\phi$ and $B_r$ are supposed to have the opposite directions between the disk upper and lower layers because these magnetic fields are generated by the entrainment and accretion of the initial $B_z$ due to the rotation (Extended Data Figure 1). Thus, the observed offset value may be the combination of the upper disk layer ($\Delta\phi_+$) and the lower disk layer ($\Delta\phi_-$).

By assuming that the radial and toroidal magnetic fields in the lower layer have the same strengths but the opposite directions to the upper layer, the angle offset of the lower layer ($\Delta\phi_-$) will be estimated as $\Delta\phi_-=\arctan\frac{-(B_r)_{\phi\perp} + (B_{z,proj})_{\phi\perp} }{-B_\phi-(B_r)_{\phi} + (B_{z,proj})_{\phi}}$. Because the 0.87 mm dust continuum emission is optically thin ($\tau<0.1$), the angular offset ($\Delta\phi$) should be the weighted average of $\Delta\phi_+$ and $\Delta\phi_-$ as 
\begin{equation}
\overline{\Delta\phi}=\frac{1}{2}\arctan\frac{\rm Stokes\  {\it U}_{\Delta\phi_+} + Stokes\  {\it U}_{\Delta\phi_-}}{\rm Stokes\  {\it Q}_{\Delta\phi_+} + Stokes\  {\it Q}_{\Delta\phi_-}}\label{eq2},
\end{equation}
where the subscripts of $\Delta\phi_+$ and $\Delta\phi_-$ in the Stokes {\it Q} and Stokes {\it U} parameters represent the emission of the upper and lower layers, respectively. The $\overline{\Delta\phi}$ value is not simply determined by the combination of $\Delta\phi_+$ and $\Delta\phi_-$  but by the combination of the Stokes {\it Q} and {\it U} emission.

The polarization fraction is calculated by $p\approx p_{\rm max}\sin^2\theta$, where $\theta$ is the projected length of the elongated dust grains along the plane of the sky \cite{1985ApJ...290..211L,2016MNRAS.460.4109Y} and $p_{\rm max}$ is defined as the intrinsic polarization, which is the maximum polarization when the grains are perfectly aligned in the same direction. However, $p_{\rm max}$ can be ignored in this calculation because we assume the same $p_{\rm max}$ between the upper and lower layers. Then, the $\overline{\Delta\phi}$ value is independent of $p_{\rm max}$. Therefore,  Stokes {\it Q} and Stokes {\it U} is only derived from $\sin^2 \theta$ in both layers.

\subsection*{Estimating the magnetic field strength and power-law of the vertical magnetic field}

From the relative strengths of the three-dimensional magnetic fields, we estimated the magnetic field strength and the power-law profile of the vertical magnetic field with radius.
By assuming the balance between the generation of $B_\phi$ by vertical shear motion and the dissipation of $B_\phi$ by ambipolar diffusion, $B_\phi$ can be estimated \cite{2023PASJ...75..835T} as 
\begin{equation}
B_\phi=\Big(\frac{H}{r}\Big)^2 \Big(\frac{B_z H}{\eta_{\rm A}}\Big) v_\phi=\frac{H^3}{r }\frac{\Omega_{\rm K}}{\eta_{\rm A}} B_z,
\end{equation}
where $H$, $r$,$\eta_{\rm A}$, $v_\phi$, $\Omega_{\rm K}$ are the gas scale height, radius, ambipolar resistivity, Kepler velocity, and Kepler frequency. Therefore, the ambipolar resistivity can be written as $\eta_{\rm A}=\frac{H^3 \Omega_{\rm K}}{r}\frac{B_z}{B_\phi}$. The gas scale height is given by $H=c_{\rm s}/\Omega_{\rm K}$ , where $c_{\rm s}$ is the sound speed. Here we assumed that the dust scale height is the same with the gas scale height because this disk is suggested to be highly turbulent \cite{2021ApJ...921..173T}. The turbulent strength ($\alpha$) and dust size ($a_{\rm dust}$) are estimated to be $\alpha\sim 0.3$ and $a_{\rm dust}\sim3$ microns, respectively, from near-infrared observations with the compact dust assumption. Then, the dust scale height can be calculated as $H_{\rm dust}=(1+\frac{{\rm St}}{\alpha}\frac{1+2{\rm St}}{1+{\rm St}})^{-1/2}  H_{\rm gas}$ \cite{1995Icar..114..237D,2007Icar..192..588Y}, where ${\rm St}=(\pi\rho a_{\rm dust})/(2\Sigma_{\rm gas})$. Therefore, $H_{\rm dust}\approx H_{\rm gas}$ when the dust size is less than 100 microns, assuming a gas surface density of 0.2 g cm$^{-2}$ and a turbulent strength $\alpha$ of 0.3. Note that the turbulent strength has been studied with different cases of porous dust grains and dust surface density by keeping a value of ${\rm St}/\alpha$ \cite{2021ApJ...921..173T}. Even if the dust structure and dust surface density are changed, our assumption of $H_{\rm dust}\approx H_{\rm gas}$ is robust as long as ${\rm St}/\alpha$ does not change.  In addition, we assume that the relative strengths of the magnetic field do not vary so much within the dust scale height. This is suggested to be a reasonable assumption for $B_z$, because the variation of $B_z$ within the gas scale height is estimated to be $\Delta B_z\approx \frac{H}{r} B_r \approx \frac{H}{r} B_z$ \cite{2011PhDT.......405B}. Here, $r\approx 200$ au and $H\approx 20$ au. Therefore, $\Delta B_z$ can be negligible. In contrast, the variations of $B_r$ and $B_\phi$ within the gas scale height are unknown parameters and have been studied by MHD simulations. If the single layer model is the case, it can be speculated that the strengths of $B_r$ and $B_\phi$ do not vary as much within the scale height \cite{2017ApJ...845...75B}. In the dual layer model, the polarized emission would be emitted from the relatively upper layer because $B_\phi$ will be dropped at the disk midplane to reverse the direction, whereas the observations indicate that $B_\phi$ is the dominant component. 
Note that the emission is estimated to be optically thin ($\tau\sim 0.1$) at a wavelength of 0.87 mm, suggesting that the polarized emission would be the averaged value along the vertical direction of the scale height, and could mainly trace the midplane region. This also supports that the single layer model is preferred to the dual layer model.

By applying the central star of $2.4$ $M_\odot$, temperature of $30$ K, and $\frac{B_z}{B_\phi}$  of 0.23, $\eta_{\rm A}$ is derived to be $\eta_{\rm A}=2.3\times10^{17}$ (cm$^2$  s$^{-1}$).  Then, because the gas density is small enough, the ambipolar resistivity is approximated as $\eta_{\rm A}=B^2/(4\pi C\gamma\rho^{3/2})$ \cite{2023PASJ...75..835T,2011MNRAS.416..591T}. Here $C$ is given as $C=\sqrt{\frac{m_i^2\zeta_{\rm CR}}{m_g\beta_r}}$, where $m_i$ and $m_g$ are the masses of the ion and neutral particles, and we adopted $m_i=29$ $m_{\rm H}$ and $m_g=2.34$ $m_{\rm H}$ assuming that the major ion is HCO$^+$, where $m_{\rm H}$ is the hydrogen mass. The HCO$^+$ molecule is suggested to be the most abundant ion when the abundance ratio of $n({\rm CO})/n({\rm electron})$ is higher than $\sim 10^3$ \cite{2015ApJ...807..120A}. Although the electron abundance is not well constrained, it is usually assumed to be $X({\rm e}) \lesssim10^{-9}$. 
On the other hand, the CO abundance is estimated to be $X({\rm CO}) \sim 10^{-4}$ for the HD 142527 disk \cite{2019PASJ...71..124S}. Therefore, we assume that the HCO$^+$ molecule is the most abundant ion for this region. $\zeta_{\rm CR}$ is the cosmic ray ionization rate, and we assume $10^{-17}$ s$^{-1}$. $\beta_r$ is the recombination rate and is assumed to be $\beta_r=\beta_{r,0} (\frac{T}{300\ \rm K})^{-0.67}$, where $\beta_{r,0}=2.4\times10^{-7}$ cm$^3$ s$^{-1}$. $\gamma=\frac{\langle\sigma v\rangle_{\rm in}}{(m_g+m_i)}$, where $\langle\sigma v\rangle_{\rm in}$ is the rate coefficient for collisional momentum transfer between ions and neutrals. We assume $\langle\sigma v\rangle_{\rm in}=1.3\times10^{-9}$ cm$^3$ s$^{-1}$. From this formula, we can estimate the magnetic field strength. The gas density $\rho$ is derived to be $\rho=\Sigma_{\rm g}/H$, where $\Sigma_{\rm g}$ is the gas surface density, which is derived to be $\Sigma_{\rm g}=0.2$ g cm$^{-2}$ \cite{2019PASJ...71..124S}. From these parameters, the magnetic field strength is finally derived to be $B \sim 0.3$ mG.
The Am parameter can also be estimated from the magnetic field strength. The Am parameter is defined as ${\rm Am}=\frac{v_{\rm A^2}}{\eta_{\rm A} \Omega_{\rm K}}=\frac{B^2}{\eta_{\rm A} \Omega_{\rm K}\times4\pi\rho}$, where $v_{\rm A}$ is the Alfv\'en velocity of $v_{\rm A}=\sqrt{\frac{B^2}{4\pi\rho}}$. By assuming $B\sim0.3$ mG, $\eta_{\rm A}=2.3\times10^{17}$ cm$^2$  s$^{-1}$, and $\rho=6.6\times10^{-16}$ g cm$^{-3}$, the Am parameter is derived to be $\rm Am \sim 0.4$. The plasma beta is defined as $\beta=\frac{8\pi\rho c_{\rm s}^2}{B^2}$ and was derived to be $\beta\sim 2.0\times10^2$.

By assuming that $B_z$ has a power-law profile of $B_z\propto r^p$, the power-law index, $p$, can be estimated from the relative strengths between $B_r$ and $B_z$. The radial magnetic field of  $B_r$ is approximately obtained as  $B_r=\frac{\Psi(r)}{2\pi r^2}$, where $\Psi(r)$ is the magnetic flux enclosed within a radius $r$ and is given by $\Psi=\int_0^r 2\pi rB_z dr$. By assuming the power-law profile of $B_z=B_{z,0} (\frac{r}{1\ \rm au})^p$, $\Psi (r)=\int_0^r 2\pi r B_{z,0} (\frac{r}{1\  \rm au})^p dr = \frac{2\pi B_{z,0}}{p+2} (\frac{r}{1 \ \rm au})^{p+2}$, where $B_{z,0}$ is the vertical magnetic field strength at a radius of 1 au. From the magnetic flux,  $B_r$ can be written as $B_r=\frac{2\pi B_{z,0}}{p+2}(\frac{r}{1 \ \rm au})^{(p+2)}  \frac{1}{2\pi r^2}=\frac{B_{z,0}}{(p+2)} (\frac{r}{1 \ \rm au})^p$. Therefore, $\frac{B_r}{B_z} =\frac{1}{(p+2)}$.
Since $\frac{|B_r |}{|B_z |} \approx 1$ around a radius of 200 au, the power-law index, $p$, is derived to be $p\approx -1$. This implies that the magnetic field increases moderately toward the center. The fact that the power exponent of the magnetic field is flatter than 2 suggests that the magnetic flux freezing has thawed and that both magnetic field advection and outward diffusion both play a role \cite{2014MNRAS.441..852G,2024PASJ...76..674T}.

\newpage

\backmatter






\section*{Declarations}


\begin{itemize}

\item Data availability 

The data sets generated and/or analyzed during the current study are available in the ALMA archive, https://almascience.nrao.edu/alma-data/archive, with program IDs \#2015.1.00425.S, \#2017.1.00987.S, \#2018.1.01172.S, and \#2022.1.00406.S.

\item Code availability 
The Python packages used in the data analysis are all publicly available. The calibration and Chi-square fitting scripts are available from the corresponding author upon reasonable request.

\item Acknowledgements

This paper makes use of the following ALMA data: ADS/JAO.ALMA\#2015.1.00425.S, \#2017.1.00987.S, \#2018.1.01172.S, and \#2022.1.00406.S. ALMA is a partnership of ESO (representing its member states), NSF (USA) and NINS (Japan), together with NRC (Canada) and NSC and ASIAA (Taiwan) and KASI (Republic of Korea), in cooperation with the Republic of Chile. The Joint ALMA Observatory is operated by ESO, AUI/NRAO and NAOJ. 
SO acknowledges support by JSPS KAKENHI Grant Numbers JP20K14533, JP22H01275, and JP23K22546. 
TM acknowledges support by JSPS KAKENHI Grant Number JP23K03463.
TT acknowledges support by JSPS KAKENHI Grant Number JP20K04017 and JP24K07097.
NS acknowledges support by JSPS KAKENHI Grant Number JP20H00182.

\item Author contribution

This project was led by S.O, who was responsible for the analysis and wrote the manuscript together with T.M., Y.T., and M.M. Y.T. provided the calculations of the physical parameters including the magnetic field strength from the relative strengths of the magnetic field. All authors contributed to the data analysis, discussed the results, and contributed to the manuscript.

\item Competing interests

The authors declare no competing interests.

\end{itemize}







\begin{appendices}






\end{appendices}


\newpage

\section{Supplementary Information}

\subsection{Imaging of the differences of the polarization vectors and polarization fractions}

To avoid confounding by different spatial frequency components, the images of the polarization vector differences (Supplementary Figure 1) were generated using the same uv distance. For $\Delta {\rm PA}_{\rm 1.3mm-0.87mm}={\rm PA}_{\rm 1.3mm} - {\rm PA}_{\rm 0.87mm}$, the Stokes {\it Q} and {\it U} images of the 0.87 mm and 1.3 mm polarization data were generated by using a uv distance of $1.4\times10^4-5.5\times10^5$ $\lambda$. The Stokes {\it Q} and {\it U} images of the 0.87 mm data were smoothed to be a beam size of $0.517''\times0.452''$ with a position angle of $64.2^\circ$ to match the 1.3 mm data. The polarization vectors were then calculated from the Stokes {\it Q} and {\it U} images where the PI emission is above the 3sigma levels. The $\sigma_{\rm PI}$ is derived to be $35$ and $23$ $\mu$Jy beam$^{-1}$ for 0.87 mm and 1.3 mm, respectively.

For $\Delta {\rm PA_{2.1mm-0.87mm}=PA_{2.1mm} - PA_{0.87mm}}$, the Stokes {\it Q} and {\it U}  images of the 0.87 mm and 1.3 mm polarization data were generated by using a uv distance of $1.4\times10^4-1.2\times10^6$ $\lambda$. The Stokes {\it Q} and {\it U}  images of the 0.87 mm data were smoothed to be a beam size of $0.479''\times0.405''$ with a position angle of $76.1^\circ$ to match the 2.1 mm data.
Then, the polarization vectors were calculated from the Stokes {\it Q} and {\it U}  images. The polarization vectors were calculated where the PI emission is above the 3sigma levels, where $\sigma_{\rm PI}$ is derived to be $33$ and $7.6$ $\mu$Jy beam$^{-1}$ at 0.87 mm and 2.1 mm, respectively.
To make the images of the ratios of the polarization fractions (Supplementary Figure 3), the polarization fractions were also calculated by the above method. The polarization fractions were calculated where the PI emission is above the 3sigma levels.

For $\Delta {\rm PA_{2.7mm-0.87mm}=PA_{2.7mm} - PA_{0.87mm}}$, the Stokes {\it Q} and {\it U}  images of the 0.87 mm and 2.7 mm polarization data were generated by using a uv distance of $1.4\times10^4-4.7\times10^5$ $\lambda$. The Stokes {\it Q} and {\it U}  images of the 0.87 mm data were smoothed to be a beam size of $0.711''\times0.678''$ with a position angle of $50.6^\circ$ to match the 2.7 mm data.
Then, the polarization vectors were calculated from the Stokes {\it Q} and {\it U}  images. The polarization vectors were calculated where the PI emission is above the 3sigma levels, where $\sigma_{\rm PI}$ is derived to be $54$ and $7.4$ $\mu$Jy beam$^{-1}$ at 0.87 mm and 2.7 mm, respectively.
To make the images of the ratios of the polarization fractions (Supplementary Figure 3), the polarization fractions were also calculated by the above method. The polarization fractions were calculated where the PI emission is above the 3sigma levels.

To evaluate the wavelength dependence of the polarization patterns, we calculated the differences of the polarization vectors at 1.3 mm, 2.1 mm, and 2.7 mm with respect to the 0.87 mm polarization vectors by deriving $\Delta {\rm PA_{1.3mm-0.87mm}=PA_{1.3mm} - PA_{0.87mm}}$, $\Delta {\rm PA_{2.1mm-0.87mm}=PA_{2.1mm} - PA_{0.87mm}}$, and  $\Delta {\rm PA_{2.7mm-0.87mm}=PA_{2.7mm} - PA_{0.87mm}}$. Supplementary Figure 1 shows the images of the differences in polarization vectors. The northern region shows $\Delta {\rm PA_{1.3mm-0.87mm}}$, $\Delta {\rm PA_{2.1mm-0.87mm}}$, and $\Delta {\rm PA_{2.7mm-0.87mm}}$ of $\sim90$ degrees on the ridge of the horseshoe structure, indicating that the $\rm PA_{0.87mm}$ is perpendicular to $\rm PA_{1.3mm}$, $\rm PA_{2.1mm}$, and $\rm PA_{2.7mm}$. In contrast, the southern region shows that the polarization vectors are almost in the same direction ($\Delta {\rm PA_{1.3mm-0.87mm}}$, $\Delta {\rm PA_{2.1mm-0.87mm}}$, and $\Delta {\rm PA_{2.7mm-0.87mm}}\sim 0$ degree). Supplementary Figure 2 plots $\Delta {\rm PA_{1.3mm-0.87mm}}$, $\Delta {\rm PA_{2.1mm-0.87mm}}$, and $\Delta {\rm PA_{2.7mm-0.87mm}}$ as a function of a position angle of $110-250$ degrees on the ridge of the southern part of the horseshoe structure. The plotted data were taken from the pixels at the ridge position with a Nyquist sampling of 11 degrees for $\Delta {\rm PA_{1.3mm-0.87mm}}$, 10 degrees for $\Delta {\rm PA_{2.1mm-0.87mm}}$, and 15 degrees for $\Delta {\rm PA_{2.7mm-0.87mm}}$.  The differences of the polarization vectors are derived to be $\Delta {\rm PA_{1.3mm-0.87mm}}=3.9\pm0.6$ degrees, $\Delta {\rm PA_{2.1mm-0.87mm}}=1\pm1$ degrees, and $\Delta {\rm PA_{2.7mm-0.87mm}}=-5.2\pm3.4$ degrees. Supplementary Figures of 1 and 2 show that the polarization vectors have almost the same directions among the wavelengths. Although the larger angle offsets are seen at longer wavelength separations, this may be due to the weaker emission at longer wavelengths. The 2.7 mm polarization emission at Band 3 is the weakest emission with the largest beam size, leading to larger uncertainties in the polarization vectors due to the noise. In addition to the polarization vectors, Supplementary Figure 3 shows the ratios of the polarization fraction $\rm P_{frac,1.3 mm}/P_{frac,0.87 mm}$, $\rm P_{frac,2.1 mm}/P_{frac,0.87 mm}$, and $\rm P_{frac,2.7 mm}/P_{frac,0.87 mm}$ (method). In the northern region, the polarization fraction is significantly different, with $\rm P_{frac,1.3 mm}/P_{frac,0.87 mm}$ reaching values as high as $\sim5$ and $\rm P_{frac,2.1 mm}/P_{frac,0.87 mm}$ and $\rm P_{frac,2.7 mm}/P_{frac,0.87 mm}$ reaching values as high as $\sim10$. In contrast, in the southern region, the polarization fraction is almost the same value as all the polarization fraction ratios show $\sim1$.

\begin{figure}[h!]
\centering
\includegraphics[width=12.cm,bb=0 0 923 601]{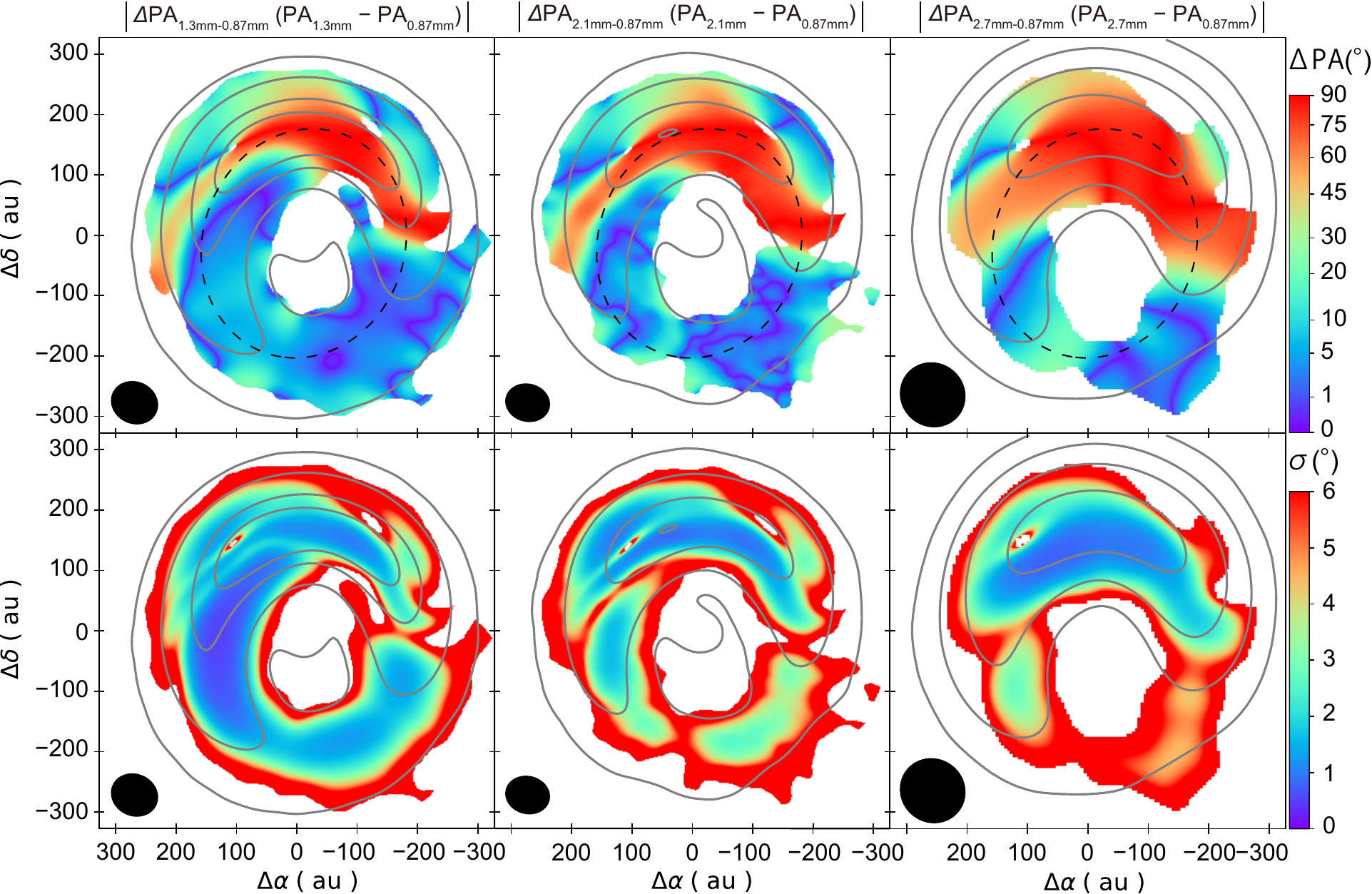}
 \captionsetup{labelformat=empty,labelsep=none}
\caption{{\bf Supplementary Figure 1 $\mid$ Images of angle differences of polarization vectors.} The angle differences of the polarization vectors, $\Delta {\rm PA}_{\rm 1.3mm-0.87mm}={\rm PA}_{\rm 1.3mm} - {\rm PA}_{\rm 0.87mm}$, $\Delta {\rm PA_{2.1mm-0.87mm}=PA_{2.1mm} - PA_{0.87mm}}$, and $\Delta {\rm PA_{2.7mm-0.87mm}=PA_{2.7mm} - PA_{0.87mm}}$ are shown.  The absolute values of the angle differences are shown in the upper panel. The errors of the values are also shown in the lower panel. The polarization vectors at 0.87 mm wavelength are perpendicular to those at 1.3 mm, 2.1 mm, and 2.7 mm wavelengths in the northern region, while they are almost in the same direction in the southern region. The contours are the brightness temperatures of the 1.3 mm emission of [0.1, 1, 5, 10, 20] K (left panels), the 2.1 mm emission of  [0.1, 1, 5, 9] K  (middle panels), and the 2.7 mm emission of [0.06, 0.3, 1, 3] K (right panels). The resolution (beam size) is shown as small ellipses in the lower left corner of each panel. The dashed lines in the top panel indicate the ridge of the ring structure derived from the 0.87 mm Stokes {\it I} image.}\label{fig5}
\end{figure}

\begin{figure}[h!]
\centering
\includegraphics[width=12.cm,bb=0 0 401 300]{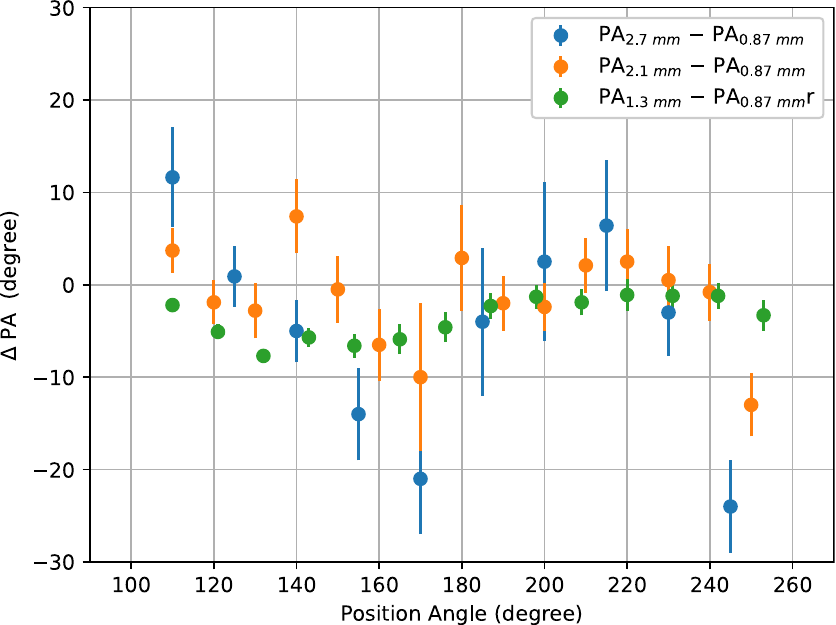}
 \captionsetup{labelformat=empty,labelsep=none}
\caption{{\bf Supplementary Figure 2 $\mid$ Plots of angle differences of polarization vectors.} The angle differences of the 1.3 mm, 2.1 mm, and 2.7 mm polarization vectors with respect to the 0.87 mm polarization vectors, $\Delta {\rm PA}_{\rm 1.3mm-0.87mm}={\rm PA}_{\rm 1.3mm} - {\rm PA}_{\rm 0.87mm}$, $\Delta {\rm PA_{2.1mm-0.87mm}=PA_{2.1mm} - PA_{0.87mm}}$, and $\Delta {\rm PA_{2.7mm-0.87mm}=PA_{2.7mm} - PA_{0.87mm}}$ as a function of a disk position angle are plotted. The positions are taken on the ridge of the horseshoe structure. The data is presented as the median of the distribution with the errors corresponding to the 68th percentiles. The differences are derived to be $\Delta {\rm PA}_{\rm 1.3mm-0.87mm}=-3.9\pm0.6$ degrees, $\Delta {\rm PA_{2.1mm-0.87mm}}=-1\pm1$ degrees, and $\Delta {\rm PA_{2.7mm-0.87mm}}=-5.2\pm3.4$ degrees suggesting that these polarization vectors are almost in the same direction within the position angle of $100-250$ degrees.}\label{fig6}
\end{figure}

\begin{figure}[h!]
\centering
\includegraphics[width=12.cm,bb=0 0 924 600]{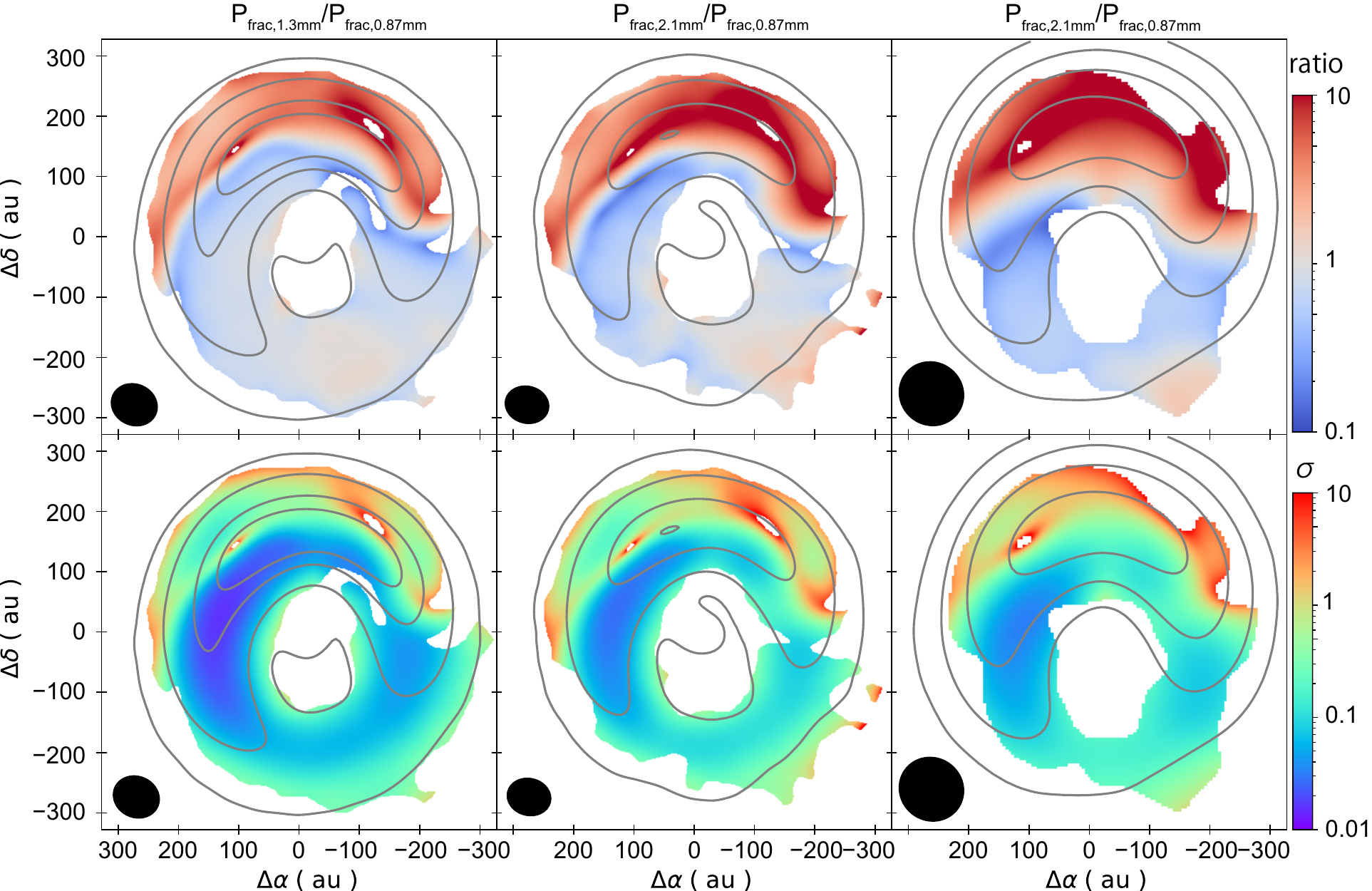}
 \captionsetup{labelformat=empty,labelsep=none}
\caption{{\bf Supplementary Figure 3 $\mid$ Ratios of the polarization fractions. } The ratios between the polarization fractions at 0.87 mm and 1.3 mm (top left panel), at 0.87 mm and 2.1 mm (top middle panel), and at 0.87 mm and 2.7 mm (top right panel). The errors of the ratios are shown in the lower panels. The northern region shows the substantial differences of the polarization fractions, while the southern region shows almost the same polarization fractions among the observed wavelengths. The beam sizes are shown in the filled ellipses in the lower left corner of each panel.}\label{fig7}
\end{figure}

 \begin{table}[h]
\rotatebox{90}{
\begin{tabular}{@{}llllllllll@{}}
\toprule
 Project ID & Antenna   & Frequency  & Wavelength  & Baseline length  & On-source Time  & Bandpass   & Gain   & Polarization   & CASA version\\
  & Number  & Band &  (mm)  & (m)  & (minutes)  &  Calibrator & Calibrator  &  Calibrator  &\\
\midrule
2015.1.00425.S    & 38 - 45   & 7  & 0.87 & 12 - 1119 & 153 & J1427-4206 & J1604-4441 & J1512-0905, J1427-4206 & 4.5.3  \\
2017.1.00987.S    & 43   & 4  & 2.1 & 15 - 2517 & 120 & J1427-4206 & J1642-4228 & J1642+3948 & 5.5.1  \\
2018.1.01172.S    & 46   & 6  & 1.3 & 15 - 704 & 60 & J1427-4206 & J1610-3958 & J1517-2422 & 5.5.1  \\
2022.1.00406.S    & 42 - 46   & 3  & 2.7 & 15 - 1397 & 600 & J1517-2422 & J1604-4441 & J1427-4206 & 6.4.1  \\
\botrule
\end{tabular}}
\captionsetup{labelformat=empty,labelsep=none}
{Supplementary Table 1: Summary of Observations}\label{tab1}%
\end{table}

\begin{table}[h]
\begin{tabular}{@{}llll@{}}
\toprule
 Band (wavelength) & robust   & Synthesized Beam  & Noise ($\mu$Jy beam$^{-1}$)  \\
\midrule
7 (0.87 mm)    & 0.5   & $0.27''\times0.24''$ ($-86.6^\circ$)  & 41 (Stokes {\it I})  \\
   &    &    & 19 (Stokes {\it Q})  \\
   &    &    & 20 (Stokes {\it U})  \\
   &    &    & 20 (PI)  \\
6 (1.3 mm)    & 0.5   & $0.51''\times0.47''$ ($-65.4^\circ$)  & 39 (Stokes {\it I})  \\
   &    &    & 23 (Stokes {\it Q})  \\
   &    &    & 23 (Stokes {\it U})  \\
   &    &    & 23 (PI)  \\
4 (2.1 mm)    & 2   & $0.48''\times0.41''$ ($-76.1^\circ$)  & 35 (Stokes {\it I})  \\
   &    &    & 8.5 (Stokes {\it Q})  \\
   &    &    & 8.5 (Stokes {\it U})  \\
   &    &    & 8.5 (PI)  \\
3 (2.7 mm)    & 2   & $0.72''\times0.68''$ ($-50.2^\circ$)  & 27 (Stokes {\it I})  \\
   &    &    & 7.4 (Stokes {\it Q})  \\
   &    &    & 7.4 (Stokes {\it U})  \\
   &    &    & 7.4 (PI)  \\
\botrule
\end{tabular}
\captionsetup{labelformat=empty,labelsep=none}
{Supplementary Table 2: Summary of Imaging parameters}\label{tab2}%
\end{table}

\end{document}